# Neural Dynamics in Parkinsonian Brain:
# The Boundary Between Synchronized and Nonsynchronized Dynamics

Choongseok Park[1], Robert M. Worth[2,1], Leonid L. Rubchinsky[1,3, *]

[1] Department of Mathematical Sciences and Center for Mathematical Biosciences, Indiana University Purdue University Indianapolis, Indianapolis, IN 46202, USA
[2] Department of Neurosurgery, Indiana University School of Medicine, Indianapolis, IN 46202, USA
[3] Stark Neurosciences Research Institute, Indiana University School of Medicine, Indianapolis, IN 46202, USA

Synchronous oscillatory dynamics is frequently observed in the human brain. We analyze the fine temporal structure of phase-locking in a realistic network model and match it with the experimental data from parkinsonian patients. We show that the experimentally observed intermittent synchrony can be generated just by moderately increased coupling strength in the basal ganglia circuits due to the lack of dopamine. Comparison of the experimental and modeling data suggest that brain activity in Parkinson's disease resides in the large boundary region between synchronized and nonsynchronized dynamics. Being on the edge of synchrony may allow for easy formation of transient neuronal assemblies.

PACS: 87.19.lm, 05.45.Xt, 87.19.ll, 87.19.xp

Synchronized oscillatory activity in the brain has long attracted the attention of physicists and can be analyzed and modeled with the approaches of nonlinear dynamics [1]. This activity is significant for motor behavior and cognition [2]. Disorganization of the strength and pattern of synchrony is related to several brain disorders [3]. In particular, brain activity in Parkinson's disease is marked by excessive synchrony of neural oscillations in the beta ($\beta$) frequency band, which not only accompanies the motor symptoms, but is likely to cause them [4]. Treatment of symptoms tends to decrease this excessive synchrony suggesting that a low level of synchrony is the healthy state. This study is aimed at understanding the dynamical nature of these oscillations from the perspective of nonlinear dynamics.

The symptoms of Parkinson's disease ultimately result from the death of dopaminergic neurons in a subcortical brain structure called the basal ganglia. For many basal ganglia synapses this will be translated into an increase of synaptic connections because dopamine tends to suppress them. For example, dopamine suppresses the release of the inhibitory neurotransmitter GABA in the Globus Pallidus pars externa (GPe) [5] and in the subthalamic nucleus (STN) [6]. This should effectively strengthen the connections between neurons. In that sense, Parkinson's disease is a naturally occurring experiment, where the parameters of the oscillatory network are varied. However, it is an uncontrolled experiment. On the other hand, *in vitro* experiments would unavoidably damage the networks, while *in vivo* experiments [4] would be hampered by a restricted set of dopaminergic states and other limitations. In a model one is not bound by these constraints. The key element of this study is matching the modeling with experimental data recorded in parkinsonian patients undergoing neurosurgical procedures. This matching allows us to elucidate the dynamical state of brain activity in Parkinson's disease and to suggest possible implications for the brain dynamics.

*Network model.* We use the conductance-based model of subthalamo-pallidal network first derived in [7] and further developed in, e.g., [8,9]. The model circuit has two chains (10 GPe and 10 STN neurons in each) with the circular boundary conditions. Each STN neuron sends a connection to one GPe neuron, and each GPe neuron sends a connection to the STN neuron from which it gets a connection and to its two nearest neighbors (Fig. 1). The architecture of the network is in agreement with basal ganglia anatomy [10]. Each neuron is described by a conductance-based, Hodgkin-Huxley-like system of differential equations. The membrane potential equation is

$$C\frac{dV}{dt} = -I_L - I_K - I_{Na} - I_T - I_{Ca} - I_{AHP} - I_{syn} + I_{app} \, , \tag{1}$$

with membrane currents: leak current $I_L = g_L(V - V_L)$, fast $K$ and $Na$ currents $I_K = g_K n^4 (V - V_K)$ and $I_{Na} = g_{Na} m_\infty^3 (V) h(V - V_{Na})$, $Ca$ currents $I_T = g_T a_\infty^3 (V) b_\infty^2 (r)(V - V_{Ca})$ and $I_{Ca} = g_{Ca} s_\infty^2 (V)(V - V_{Ca})$, and $Ca$-activated voltage-independent $K$-current $I_{AHP} = g_{AHP}([Ca]/([Ca]+k_1))(V - V_K)$. $m_\infty$, $a_\infty$ and $s_\infty$ are instantaneous voltage-dependent gating variables (they do not depend on time explicitly); $b_\infty$ is sigmoidal function of time dependent variable $r$ (described by Eq. 3, [7]). The concentration of intracellular $Ca^{2+}$ is governed by the calcium balance equation

$$d[Ca]/dt = \varepsilon(-I_{Ca} - I_T - k_{Ca}[Ca]) \, . \tag{2}$$

$n, h$ and $r$ are gating variables obeying

$$dx/dt = (x_\infty(V) - x)/\tau(V) \, . \tag{3}$$

Thus each model neuron is a 5-dim system with variables $V, n, h, r$, and $[Ca]$. Connections between network elements (excitatory and inhibitory synapses) are modeled by an equation for the fraction of activated channels

$$\frac{ds}{dt} = \alpha H_\infty(V_{presyn} - \theta_g)(1-s) - \beta s \, , \tag{4}$$

with $H_\infty(V) = 1/(1 + \exp[-(V - \Theta_g^H)/\sigma_g^H])$. Synaptic current $I_{syn} = g_{syn}(V - V_{syn})\sum_j s_j$, where summation is over *s*-variables from all neurons projecting to a given neuron. See [11] for model details and parameter values. Thus our modeling retains the neuronal properties and basic anatomy of GPe-STN networks. Twenty neurons in this network is sufficiently large number to generate patterns of activity, relevant to brain dynamics [7].

The model equations describe electrical activity of STN neurons for which experimental data are available (see below). In addition, the experimental data also includes local field potentials (LFP), a low-frequency signal generated by the synaptic potentials and thus reflecting incoming and local processing activity [12]. In particular, STN LFP has synaptic origin and is locally generated [13]. There are no connections between STN neurons [14], thus STN LFP has its origin in pallido-subthalamic synaptic transmission. Let $I_{GS}^i = g_{GS}(V^i - V_{syn})\sum_{j=i-1}^{i+1} s_j$ be synaptic input to the $i^{th}$ STN neuron. Then the model STN LFP is $LFP^i = I_{GS}^i + w(I_{GS}^{i-1} + I_{GS}^{i+1})$, where *w* is the weight, representing the impact of more remote synaptic activity. Inclusion of an additional term $sw(I_{GS}^{i-2} + I_{GS}^{i+2})$ with small *sw* did not substantially change the results.

*Technique to study the dynamics of phase-locking.* The modeling was compared with the same experimental data as presented in [15]: a spiking signal and an LFP recorded in the STN of 8 parkinsonian patients. The dynamics of the model was analyzed with the time-series analysis techniques identical to [15]. They are based on the analysis of phase synchronization [16] on short time-scales [17]. Briefly, both spikes and LFPs were filtered to the β-band, defined here as 10-30Hz range. The Hilbert phase was computed and first-return maps were constructed: whenever the phase of the LFP signal crossed a specified check point from negative to positive values, we recorded the spiking signal phase, generating a set of consecutive phase values $\{\varphi_{spikes,k}\}, k = 1,...N$. Thus $\varphi_k$ represents the phase difference between the spiking unit and LFP activity. Then $\varphi_{spikes,k+1}$ vs. $\varphi_{spikes,k}$ was plotted for $k = 1,...N-1$. Fig. 2A has a diagram for the first-return map. The rates $r_i$ of transitions between four regions of the map are defined as the number of points (in a given region), from which the system evolves along the arrow, divided by the total number of points in that region [15]. This approach allows for comparison of not only overall degree of phase-locking, but also of the timing of synchronized/desynchronized intervals. Moreover, the rates $r_{2,3,4}$ characterize the parts of the phase space away from the synchronized state. This ensures more adequate match of experiment and model.

*Dynamics of the model network vs. experimental data.* Our recent experimental study of oscillatory activity in parkinsonian patients revealed specific temporal structure of the β -band synchronization [15]. During episodes with some overall synchrony present, there is a predominance of short but numerous events when phase-locking of the spiking signal and the LFP is lost. The examples of the first-return maps for the experiment and the model are in Fig. 2B,C. Similar to the experiment, the model dynamics spends most of the time in the region I, which is termed the synchronized state (the phase of the spiking signal with respect to the phase of the LFP is not changed much from a cycle to cycle). For the set of parameter values used at this particular simulation, the model system exhibits the same temporal patterns of synchronization as the parkinsonian brain does. More quantitatively, the transition rates from the model are close to the experimental transition rates (Fig. 3A). The distributions of durations of desynchronization events in the model and in the experiment are similar too (Fig. 3B). Thus, the mechanisms and assumptions of the model network (bursting properties and GPe-STN connectivity) are sufficient to generate the intermittency of synchrony observed in the experiments.

We further varied model parameters to identify the parameter domain where the model dynamics is similar to the experimental dynamics. We studied the model dynamics under variation of two parameters – synaptic strength of GPe-STN projections $g_{syn}$ and the applied current term in the GPe model neuron $I_{app}$ (incorporates the effect of striatopallidal synapses). As we mentioned earlier, both synapses are functionally suppressed by the action of nigral dopamine, which is lacking in Parkinson's disease [18]. As a criterion of similarity between model and experiment we required all four rates $r_i$ in the model to be within 0.7SD of the experimental rates (given at the Fig. 3A). Since the durations of desynchronized episodes are closely related to these rates in the data [15] this requirement will also lead to similarity of the durations.

The results of this numerical experiment are summarized at Fig. 4. The right lower corner of this diagram (higher $g_{syn}$ and lower $I_{app}$) corresponds to lower dopamine levels (extreme parkinsonian state) while the higher dopamine levels (more healthy state) are closer to the upper left corner. Fig. 4 indicates that the right lower corner of the parameter plane is a region of much correlated activity (there are a small number of principle components) while the left upper corner exhibits uncorrelated dynamics. The region where the model activity is "dynamically" similar to the activity in Parkinson's disease is situated in between these extremes. Its exact location and form depend on how strong the similarity requirements are and how the model LFPs are computed, but qualitatively these do not change its location. We varied the parameter, *w,* used in the model LFP computation over a relatively broad range and the results do not exhibit strong qualitative dependence on the value of this weight (see Fig. 4). Thus the location of this region of parkinsonian synchrony is robust and is on the boundary between synchronized and nonsynchronized dynamics in the network. What adds to the confidence in the comparison of experimental data and model is that the comparison of all four rates $r_{1,2,3,4}$ described the similarity of model and experimental phase spaces not only in a vicinity of synchronized state, but also away from it (rates $r_{2,3,4}$). So in terms of the analysis of intermittency, not only the model described the "laminar" (synchronized) phase, but it also described "non-laminar" (desynchronized) phase.

*Implications of the modeling results.* There may be many more dopamine-modulated parameters than we consider here. However, if the lack of dopamine leads to stronger functional connections, we expect the results do not change much. While our study does not exclude other mechanisms contributing to the intermittent nature of synchrony in the real data (e.g., noise, fluctuating sensory inputs or short-term synaptic plasticity), it suggests that the network mechanisms by themselves are capable of intermittent synchrony generation.

Transitions between synchronized and non-synchronized dynamics has being studied theoretically, in particular, in the relatively simple models of cortical networks such as [19], which allow some analytical treatment. However, those studies did not provide a quantitative comparison of the models with experiment, which is at the core of this letter. Our comparison of modeling and experimental data indicates that at rest in Parkinson's disease the oscillations are at the boundary of synchronous and non-synchronous regimes.

There are experimental examples of transient synchrony in the brain, needed for physiologically significant events [2,20] and transient dynamics in the nervous system has being suggested to be generic [21]. In these examples transient synchrony is recruited to achieve a particular physiological effect. In our case, in Parkinson's disease even at rest the synchrony is easy to form. Interestingly, the dynamics of synchronous oscillations in the rest tremor in Parkinson's disease bears an intermittent character too [22].

Synchronous oscillations in the β-band are related to the preparation of a new movement or maintenance of the current motor set [23]. Moving further away from the boundary region into the region of nonsynchronized dynamics, the network may enter a supposedly healthy state. In that state synchrony may be possible only if other parts of neural system (transiently) bring it close to the boundary. The results may allow one to hypothesize that the healthy network is functionally close to the boundary area too, so that synchrony of β -band oscillations may be generated when needed (to prepare for a new movement or to maintain motor status quo). In a pathological state, the coupling in the network is stronger due to the lack of dopamine. This moves the network towards the more synchronized state and the transient synchronous patterns become more prevalent and harder to break, which would prevent the execution of the new movement.

Our results call for further study of intermittent synchrony in the neural systems. Studies of stochastic variants of the models will be able to clarify the potential action of noise on the synchrony. A similar modeling approach to the data from other brain regions where the synchronization is known to be functionally important (such as many perceptual and cognitive phenomena) and, eventually, direct experiments will clarify how generic the intermittent synchrony really is in living networks.

Our study may also be useful in studying suppression of pathological synchrony via adaptive deep brain stimulation, aimed at making the synchronized state unstable [24]. However, the real pathological state is not a stable synchrony, but intermittent synchronized dynamics. So the stimulation would act on a dynamics, which is not fully synchronous to begin with. Thus one may need to further explore the action of desynchronizing stimulation on dynamics which are not fully synchronized.

*Conclusions.* Variability of synchronous oscillations may be generated by brain networks due to a moderate increase in coupling strength, which is expected to be induced by the lack of dopamine in Parkinson's disease. Comparison of experimental and modeling data suggest that synchronized oscillations in the parkinsonian brain are on the boundary between synchronized and uncorrelated dynamics. Operation at the edge of synchrony would yield creation and disappearance of unstable synchronized clusters possible without much expense, allowing for easy formation and dissolution of transient neuronal assemblies. In turn, this would either directly contribute to a fluid and timely sequence of movements or facilitate information transmission through the oscillations [2]. The model also may be useful to study control schemes for suppressing the synchronous activity in Parkinson's disease.

This study was supported by NIH grant R01NS067200 (NSF/NIH CRCNS).

* Corresponding author: leo@math.iupui.edu
1. M. I. Rabinovich et al., Rev. Mod. Phys. **78**, 1213 (2006); T. Nowotny, R. Huerta and M. I. Rabinovich, Chaos **18**, 037119 (2008).
2. J. N. Sanes and J. P. Donoghue, Proc. Natl. Acad. Sci. U.S.A. **90**, 4470 (1993); A. K. Engel, P. Fries, and W. Singer, Nat. Rev. Neurosci. **2,** 704 (2001); G. Buzsáki, and A. Draguhn, Science **304**, 1926 (2004).
3. A. Schnitzler, and J. Gross, Nat. Rev. Neurosci. **6**, 285 (2005); P. J. Uhlhaas, and W. Singer, Neuron **52**, 155 (2006).
4. W. D. Hutchison *et al*., J. Neurosci. **24**, 9240 (2004); C. Hammond, H. Bergman, and P. Brown, Trends. Neurosci. **30**, 357 (2007); P. Brown, Curr. Opin. Neurobiol. **17**, 656 (2007); A. Eusebio and P. Brown, Exp. Neurol. **217**, 1 (2009).
5. M. Ogura, and H. Kita, J. Neurophysiol. **83**, 3366 (2000); A. J. Cooper, and I. M. Stanford, Neuropharmacology. **41**, 62 (2001).
6. K. Z. Shen, and S. W. Johnson, Neuroreport **16**, 171 (2005); J. Baufreton, and M. D. Bevan, J. Physiol. **586**, 2121 (2008).
7. D. Terman *et al*., J. Neurosci. **22,** 2963 (2002).
8. L. L. Rubchinsky, N. Kopell, and K. A. Sigvardt, Proc. Natl. Acad. Sci. U.S.A. **100,** 14427 (2003); J. E. Rubin, and D. Terman, J. Comput. Neurosci. **16**, 211 (2004).
9. J. Best *et al*., J. Comput. Neurosci. **23**, 217 (2007).
10. C. J. Wilson, Basal Ganglia. In: Shepherd GM, editor. *The Synaptic Organization of the Brain.* pp. 361-413. (Oxford University Press, New York, 2004).
11. Details of the neuronal models are available at [7,9]. Few parameter values from [9] were modified to get more realistic firing rates (the firing rates in humans are higher than those at [7], which may be due to at least three factors: *in vitro* data, rodent tissue, and low (room) temperature used for model development). In STN: $g_{Na} = 50 nS/\mu m^2$, $g_K = 40 nS/\mu m^2$, $\phi_n = \phi_h = 5$, $\phi_r = 2$, $I_{app} = 10 pA/\mu m^2$; in GPe: $K_{Ca} = 3$, $\epsilon = 0.0055$, $\phi_n = 0.3, \phi_h = 0.1$. Synaptic parameters follow [7] except $g_{STN \to GPe} = 0.4$, $\beta_{GPe} = 0.14$.
12. G. Buzsaki, R. D. Traub, and T. A. Pedley, The cellular basis of EEG activity. In: Ebersole JS, Pedley TA, editors. *Current Practice of Clinical Electroencephalography.* pp. 1-11. (Lippincott Williams & Wilkins, Philadelphia, 2003); C. Bédard, H. Kröger, and A. Destexhe, Phys. Rev. E **73**, 051911 (2006)
13. J. A. Goldberg *et al*., J. Neurosci. **24**, 6003 (2004); P. J. Magill *et al*., J. Neurophysiol. **92**, 700 (2004); P. Brown and D. Williams, Clin. Neurophysiol. **116**, 2510 (2005).
14. C. L. Wilson, M. Puntis, and M. G. Lacey, Neuroscience **123**, 187 (2004).
15. C. Park, R. M. Worth, and L. L. Rubchinsky, J. Neurophysiol. **103**, 2707 (2010).
16. A. Pikovsky, M. Rosenblum, and J. Kurths, *Synchronization: A Universal Concept in Nonlinear Sciences* (Cambridge University Press, Cambridge, UK, 2001).
17. J. M. Hurtado, L. L. Rubchinsky, and K. A. Sigvardt, J. Neurophysiol. **91**, 1883 (2004).
18. Note that $I_{app}$ is positive in the model (as it was developed in [7]), so decrease of pallidal excitability due to increase of striatal inhibition is represented by the decrease in $I_{app}$.

19. N. Brunel, J. Comput. Neurosi. **8**, 183 (2000); D. Hansel and G. Mato, Neural Comput. **15**, 1 (2003).
20. T. Kenet *et al*., Nature **425**, 954 (2003); D. Gervasoni *et al*., J. Neurosci. **24**, 11137 (2004).
21. M. Rabinovich, R. Huerta, and G. Laurent, Science **321**, 48 (2008); I. Tsuda, Chaos **19**, 015113 (2009).
22. J. M. Hurtado *et al.,* J. Neurophysiol. **93**, 1569 (2005).
23. A. K. Engel and P. Fries, Curr. Opin. Neurobiol. **20**, 156 (2010).
24. O. V. Popovych, C. Hauptmann, and P. A. Tass, Phys. Rev. Lett. **94**, 164102 (2005); N. Tukhlina et al., Phys. Rev. E **75**, 011918 (2007).

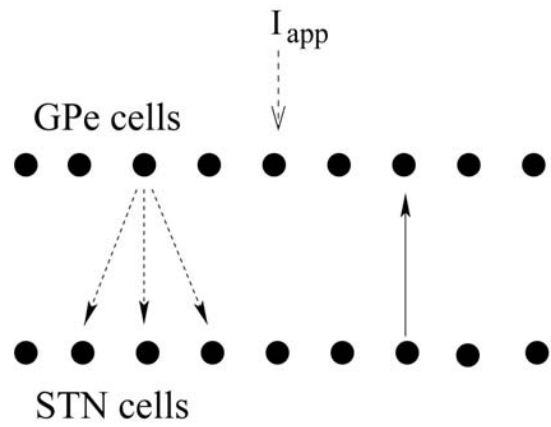

**FIG 1.**

Model network. Arrows indicate connections between cells (solid – excitatory, dashed – inhibitory).

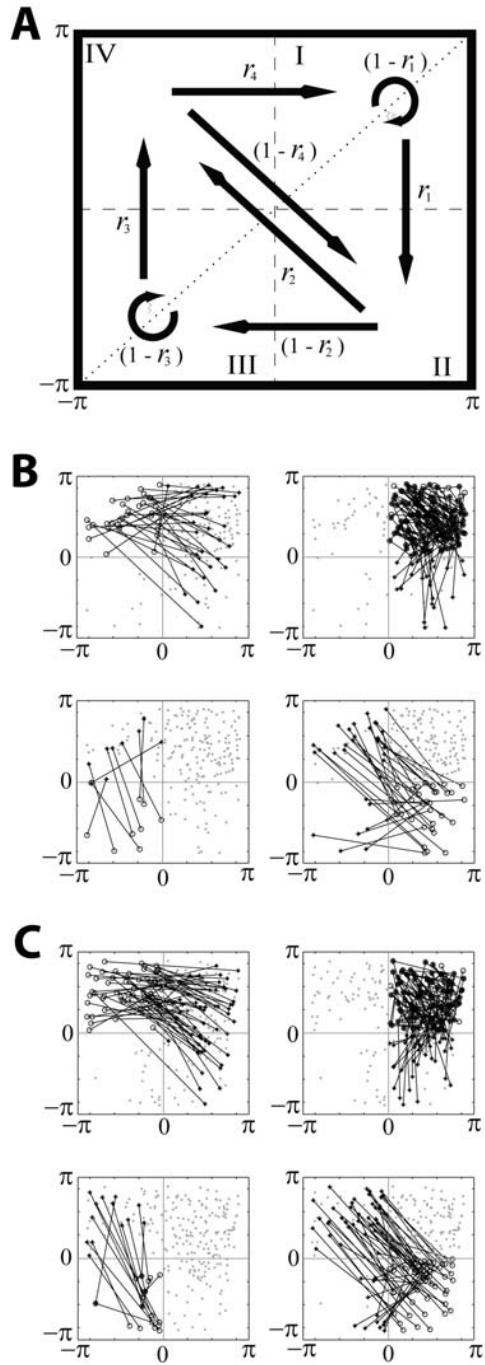

**FIG 2.**

First-return maps. A is the diagram of the map. The phase space is partitioned into four regions; arrows indicate all possible transitions between them, transition rates are written next to the arrows. B and C present examples of the maps derived from experimental and model data respectively. Four maps in B and C have the same data points, lines illustrates the transitions from points in different regions.

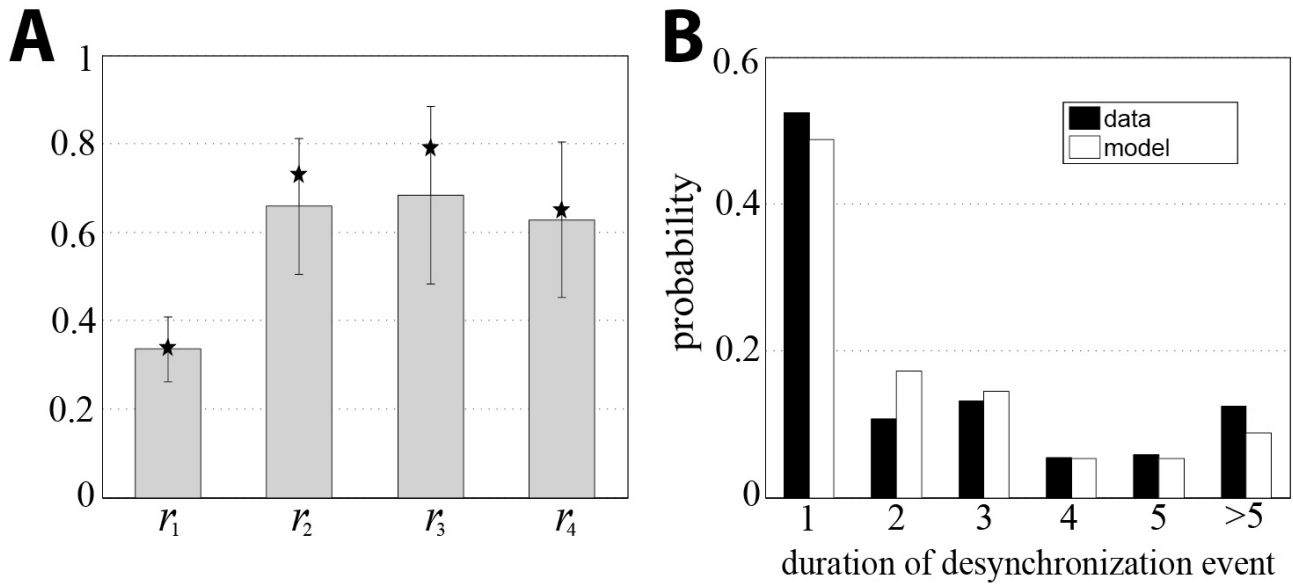

**FIG 3.**

Experimental and modeling transition rates (A) and duration of desynchronization events (B). Stars represent modeling rates, grey bars and standard deviations are for the experiment. Duration is measured in the number of cycles of oscillations. The distributions are similar and share the same features (dominance of short desynchronization events). Parameter values are the same as in Fig. 2C, experimental rates are from [15].

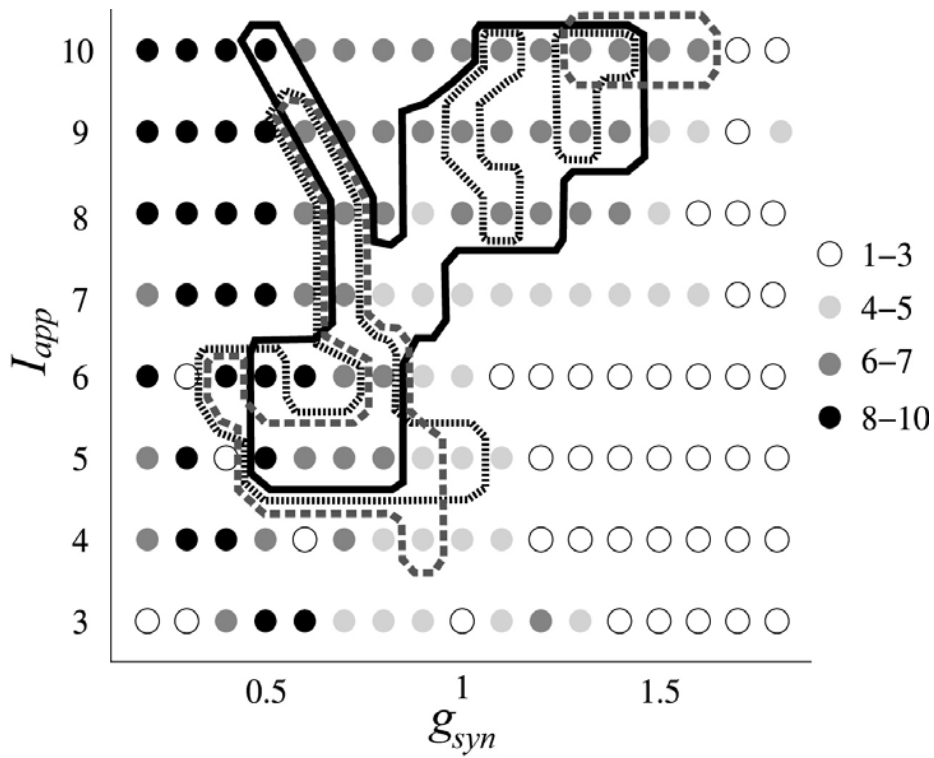

**FIG 4.**

The model dynamics in dependence on $g_{syn}$ and $I_{app}$ vs. experimental data. The circles indicate the number of principle components capturing 80% of variation in the principle component analysis for the slow variables $r$ of all STN cells in the model. The lines represent the contours of the parameter domains, where the model rates $r_i$ characterizing the temporal structure of synchronization are within 0.7SD of the experimental rates. Solid, dotted and dashed lines correspond to different values of the weights used to compute model LFP (0.26, 0.3, and 0.35).